\documentclass[apj]{emulateapj}
\usepackage{txfonts}

\begin{document}

\journalinfo{{\sc The Astrophysical Journal},
000, \pageref{start}--\pageref{end}, {\rm May 10}}
\slugcomment
{to appear in the Astrophysical Journal (vol. 642 no. 2; 2006 May 10)}
\shorttitle{CUSP SLOPE--CENTRAL ANISOTROPY THEOREM}
\shortauthors{AN \& EVANS}

\title{A Cusp Slope--Central Anisotropy Theorem}

\author{Jin H. An\altaffilmark{1,2} and N. Wyn Evans\altaffilmark{1}
\\{\footnotesize\tt jinan@space.mit.edu, nwe@ast.cam.ac.uk}}
\altaffiltext{1}{Institute of Astronomy, University of Cambridge,
Madingley Road, Cambridge, CB3 0HA, UK}
\altaffiltext{2}{MIT Kavli Institute for Astrophysics \& Space Research,
Massachusetts Institute of Technology,
77 Massachusetts Ave., Cambridge, MA 02139}

\begin{abstract}
For a wide class of self-gravitating systems, we show that if the
density is cusped like $r^{-\gamma}$ near the center, then the
limiting value of the anisotropy parameter $\beta=1-\langle
v_\mathrm{T}^2 \rangle / (2\langle v_r^2 \rangle)$ at the center may
not be greater than $\gamma/2$. Here $\langle v_r^2 \rangle$ and
$\langle v_\mathrm{T}^2 \rangle$ are the radial and tangential
velocity second moments. This follows from the nonnegativity of the
phase-space density. We compare this theorem to other proposed
relations between the cusp slope and the central anisotropy to clarify
their applicabilities and underlying assumptions. The extension of
this theorem to tracer populations in an externally imposed potential
is also derived. In particular, for stars moving in the vicinity of a
central black hole, this reduces to $\gamma\ge\beta+\case12$, indicating
that an isotropic system in Keplerian potential should be cusped at
least as steep as $r^{-1/2}$.  Similar limits have been noticed before
for specific forms of the distribution function, but here we establish
this as a general result.
\end{abstract}


\section{Introduction}

Numerical simulations of halo formation provide strong evidence that
the inner parts of dark matter halos are strongly cusped. Typically,
the density profile $\rho$ behaves like $r^{-\gamma}$, where $\gamma$
lies between 1 and 1.5 \citep*{NFW95,Mo98}. Although this numerical
result seems well established, observational evidence that dark halos
are cusped has been surprisingly elusive. A disparate body of data --
including the rotation curves of dwarf spiral galaxies \citep{PW00,dB01},
the kinematics of Local Group dwarf spheroidal galaxies \citep{KW03}, and
mass models of gravitational lens systems \citep*{TKA98} -- seem to favor
constant-density cores.

Recently, \citet{Ha04} claimed that the only density slopes permitted
by the spherically symmetric and isotropic Jeans equations are $1\le
\gamma\le3$ if the phase-space density-like quantity, $\rho / \langle
v^2 \rangle^{3/2}$, follows a power law \citep{TN01}. This result was
inferred from the condition that the power-law solution of the Jeans
and Poisson equations is physical, subject to the ``equation of state''
(EOS), $\rho\propto r^{-p}{\langle v_r^2 \rangle}^{c}$, where $p$ and
$c$ are constants. He further argued that, for the system with an
anisotropic velocity dispersion tensor, $1+\beta\le\gamma\le3$. Here
the anisotropy parameter is $\beta=1-\langle v_\mathrm{T}^2 \rangle /
(2\langle v_r^2 \rangle)$, where $\langle v_r^2 \rangle$ and $\langle
v_\mathrm{T}^2 \rangle$ are the radial and the tangential velocity
second moments \citep{BT87}. Given the nearly isotropic conditions
found in the central parts of simulated dark halos, this already seems to
indicate that the density profile cannot be shallower than
$\rho\sim r^{-1}$.

The idea of looking for constraints between the central density slope
and the anisotropy is an excellent one. In \S~\ref{sec:rel}, we show
that the inequality $\gamma\ge2\beta$ is a necessary condition for the
nonnegativity of the distribution function (DF). This generalizes two
well-established results: (1) a spherical system with a hole in
the center cannot be supported by an isotropic velocity dispersion
tensor \citep[i.e., $\gamma\ge2\beta=0$; see][]{Tr94}, and (2) a
spherical system with particles in purely radial orbits cannot be
supported by a density cusp shallower than the isothermal cusp
\citep[i.e., $\gamma\ge2\beta=2$; see][]{RT84}. In
\S~\ref{sec:hansen}, we study the scale-free power-law cusps that
inspired \citet{Ha04} and show that the inequality $\gamma\ge1+\beta$
is related to the boundary condition at infinity rather than at the
center. Finally, in \S~\ref{sec:tim}, we consider the generalization
of the theorem to tracer populations in an externally imposed cusped
potential -- for example, to stars moving around a central black
hole. In particular, with the potential cusped as $r^{-\delta}$
($0\le\delta\le1$), we find that the inequality becomes
$\gamma\ge2\beta+(\case12-\beta)\delta$.  While a similar limit has been
noted in the literature for the case in which the DF is scale-free
\citep*{Wh81,BMZ}, we derive the limit as a general condition for a DF
with a centrally diverging (and not necessarily self-consistent)
potential.

\label{start}
\section{Proof of a Cusp Slope-Anisotropy Theorem}
\label{sec:rel}

Here we shall prove that if the density is cusped like $r^{-\gamma}$
near the center, then the limiting value of $\beta$ at the center may
not be greater than $\gamma/2$. This relation is indeed suggested by
solving the Jeans equation for constant $\beta$. That is, the
one-dimensional radial velocity dispersion obtained as the solution,
in general, diverges at the center if $\gamma<2\beta$, which is
unphysical unless the central potential well depth is infinite.
In fact, we will show that the inequality $\gamma\ge2\beta$ is the
necessary condition for the nonnegativity of the DF.

\subsection{Constant Anisotropy Distribution Functions}
\label{sec:con}

First, let us suppose that the DF is given by the
\textit{Ansatz}\footnote{Strictly speaking, the \textit{Ansatz}
is valid only for
$\beta<1$. However, it can be extended to $\beta=1$ using the relation
$\lim_{\beta\rightarrow1}[L^{2\beta}/\Gamma(1-\beta)]=\delta(L^2)$,
where $\delta(x)$ is the Dirac delta ``function.'' Subsequently,
equations (\ref{eq:den})--(\ref{eq:disintc}) are still valid for
$f(E,L)=\delta(L^2)f_E(E)$ as the $\beta=1$ limit simply without the
$\Gamma(1-\beta)$ factor.}
\begin{equation}
f(E,L) = L^{-2\beta} f_E(E),
\label{eq:ansatz}
\end{equation}
where $L$ is the specific angular momentum, and $f_E(E)$ is a function
of the binding energy $E$ alone. This seems as though it is a
restrictive assumption, but this is not really the case. Rather,
equation~(\ref{eq:ansatz}) arises naturally as the most divergent term
in a Laurent series expansion with respect to $L$ at $L = 0$ for a very wide
class of DFs. By integrating equation~(\ref{eq:ansatz}) over velocity
space, we find that the density is given by
\begin{equation}
\rho = r^{-2\beta}D_\beta
\int_0^\psi\! (\psi-E)^{1/2-\beta} f_E(E)\, dE,
\label{eq:den}
\end{equation}
\begin{equation}
D_\beta=\frac{(2\pi)^{3/2}\Gamma(1-\beta)}{2^\beta\Gamma(3/2-\beta)},
\end{equation}
where $\psi$ is the relative potential and $\Gamma(x)$ is the gamma
function. The unknown function $f_E(E)$ can be found from the formula
\citep{Cu91,Ko96,WE99,AE06,EA06}
\begin{equation}
f_E(E) = C_\beta \left[
\int_0^E \frac{d\psi}{(E-\psi)^\alpha} \frac{d^{n+1}h}{d\psi^{n+1}}
+ \frac{1}{E^\alpha} \left.\frac{d^nh}{d\psi^n}\right|_{\psi=0} \right]
\label{eq:disint}
\end{equation}
\begin{equation}
C_\beta=\frac{2^\beta}{(2\pi)^{3/2}\Gamma(1-\alpha)\Gamma(1-\beta)}.
\label{eq:disintc}
\end{equation}
Here $h=r^{2\beta}\rho$ is expressed as a function of $\psi$, and
$n=\lfloor\case32-\beta\rfloor$ and $\alpha=\case32-\beta-n$ are the integer
floor and the fractional part of $\case32-\beta$. It is a simple exercise
to show that the anisotropy parameter for this model is the same as
the parameter $\beta$ in the expression of DF (eq.~\ref{eq:ansatz}).

By considering the limit of equation~(\ref{eq:den}) as $r\rightarrow0$
one can infer that $\rho$ should diverge at least as fast as
$r^{-2\beta}$ if $\beta>0$ and cannot approach zero faster than
$r^{2(-\beta)}$ if $\beta<0$ unless the integral vanishes in the same
limit. This in fact is the basic argument that leads to the theorem.
In the following, we provide a more stringent proof of the theorem.

\subsubsection{The Case $\beta \le \case12$}

For $\beta<\case12$, a direct generalization of the proof of \citet{Tr94}
for isotropic models suffices. That is, from equation~(\ref{eq:den}),
we find that
\begin{equation}
\frac{d}{d\psi}\left(r^{2\beta}\rho\right)=\tilde D_\beta
\int_0^\psi\!\frac{f_E(E)\,dE}{(\psi-E)^{1/2+\beta}}\ge0,
\label{eq:dhp}
\end{equation}
\begin{equation}
\tilde D_\beta=\left(\frac12-\beta\right)D_\beta=
\frac{(2\pi)^{3/2}\Gamma(1-\beta)}{2^\beta\Gamma(1/2-\beta)}
\end{equation}
for any physical DF, as the integrand is always positive. Similarly,
if $\beta=\case12$, then
\begin{equation}
\frac{d}{d\psi}\left(r\rho\right)=
2\pi^2\frac{d}{d\psi}\int_0^\psi\!f_E(E)\,dE=2\pi^2f_E(\psi)\ge0.
\end{equation}
However,
\begin{equation}
\frac{dh}{d\psi}=\frac{dr}{d\psi}\frac{h}{r}\frac{d\ln{h}}{d\ln{r}}
=\frac{r^{2\beta-1}\rho}{d\psi/dr}
\left(2\beta+\frac{d\ln{\rho}}{d\ln{r}}\right).
\label{eq:dhdp}
\end{equation}
Since $d\psi/dr \le 0$ for any physical potential, we thus find that
\begin{equation}
2\beta\le-\frac{d\ln{\rho}}{d\ln{r}}.
\end{equation}
This holds everywhere. Specializing to the limit at the center, we
obtain the desired result.

\subsubsection{The Case $\beta > \case12$}
\label{sec:simon}

When $\beta > \case12$, equation~(\ref{eq:dhp}) is invalid, and therefore
a different proof is required. For this purpose, we first note that
\begin{equation}
\frac{1}{(\psi-E)^{\beta-1/2}}>\frac{1}{\psi^{\beta-1/2}}>0
\end{equation}
for $0<E<\psi$ and $\beta>\case12$. Then from equation~(\ref{eq:den}),
we find that
\begin{equation}
\rho>\frac{D_\beta}{r^{2\beta}\psi^{\beta-1/2}}\int_0^\psi\!f_E(E)\,dE.
\end{equation}
Therefore,
\begin{equation}
\lim_{r\rightarrow0}\left(\rho r^{2\beta}\psi^{\beta-1/2}\right)
\ge D_\beta\int_0^{\psi_0}\!f_E(E)\,dE>0,
\label{eq:lim}
\end{equation}
where $\psi_0=\psi(r=0)$. Next, if we consider the case in which
$\psi_0$ is finite (see \S~\ref{sec:tim} for the centrally diverging
potential), we have
\begin{equation}
\lim_{r\rightarrow0}\left(\rho r^{2\beta}\psi^{\beta-1/2}\right)
=\psi_0^{\beta-1/2}\lim_{r\rightarrow0}\left(\rho r^{2\beta}\right)>0.
\end{equation}
Since $\psi_0>0$, we find that $\lim_{r\rightarrow0}(\rho
r^{2\beta})>0$, that is, $\lim_{r\rightarrow0}h$ either is nonzero
positive and finite or diverges to positive infinity.  If the former
is the case, it is straightforward to show that (using l'H\^opital's
rule)
\begin{equation}
\lim_{r\rightarrow0}h=\lim_{r\rightarrow0}\frac{hr}{r}
=\lim_{r\rightarrow0}\frac{d(hr)}{dr}
=\lim_{r\rightarrow0}h\frac{d\ln(hr)}{d\ln r},
\end{equation}
and consequently that
\begin{equation}
\lim_{r\rightarrow0}\frac{d\ln(hr)}{d\ln r}=1\qquad
\Rightarrow\
\lim_{r\rightarrow0}\frac{d\ln h}{d\ln r}=0.
\end{equation}
On the other hand, if $\lim_{r\rightarrow0}h$ is divergent,
l'H\^opital's rule indicates that
\begin{equation}
\lim_{r\rightarrow0}\frac{d\ln h}{d\ln r}
=\lim_{r\rightarrow0}\frac{\ln h}{\ln r}.
\end{equation}
However, we have $(\ln h/\ln r)<0$ for sufficiently small $r$, so
\begin{equation}
\lim_{r\rightarrow0}\frac{\ln h}{\ln r}\le0\qquad
\Rightarrow\
\lim_{r\rightarrow0}\frac{d\ln h}{d\ln r}\le0.
\end{equation}
Hence, for both cases,
\begin{equation}
\lim_{r\rightarrow0}\frac{d\ln h}{d\ln r}
=\lim_{r\rightarrow0}\frac{d\ln\rho}{d\ln r}+2\beta\le0
\end{equation}
\begin{equation}
\Longrightarrow\
\gamma=-\lim_{r\rightarrow0}\frac{d\ln\rho}{d\ln r}\ge2\beta,
\label{eq:gamma}
\end{equation}
which is the desired result.

\subsection{General (Analytic) Distribution Functions}
\label{sec:laur}

This result is in fact far more general than the assumed \textit{Ansatz}
for the DF (eq.~\ref{eq:ansatz}). For example, \citet{CP92} found
the same limit ($\gamma\ge2\beta=0$)
for the \citet{Os79,Os79tr}-\citet{Me85} type DF,
which is manifestly not in the form of equation~(\ref{eq:ansatz}).
We extend the limit derived in the preceding section
to a wide class of DFs by the following simple argument.
In general, any analytic DF can be expressed either
as sums of equation~(\ref{eq:ansatz}) or in terms of a Laurent series
expansion with respect to $L$ at $L=0$ (really a special class of the
former). Then since $L=rv_\mathrm{T}$, as $r\rightarrow0$, the DF is
dominated by the term associated with the leading order of $L$, and
consequently so is the behavior of the density near the center. It is
also straightforward to show that the anisotropy parameter at the
center is indeed determined by $\beta_0$, where `$-2\beta_0$' is the
power to the leading term of $L$. The desired result therefore
follows from the preceding proof for the special case of the DF with a
single term. In Appendix, we discuss the conditions of
applicability of the proof more mathematically and argue that the
theorem holds for all physically reasonable DFs of spherical systems.

We find that the cusp slope--anisotropy theorem ($\gamma\ge2\beta$) is
actually quite reasonable. It implies that, if the anisotropy is
radially biased ($\beta>0$) near the center, the density is cusped,
and that, unless the cusp slope is steeper than that of isothermal
cusp ($\gamma=2$), there is a finite upper limit to $\beta$ that is
strictly smaller than unity. Similarly, if the density is cored
($\gamma=0$), the central anisotropy is either tangentially biased or
at most isotropic.

\section{Scale-free Density Profile}
\label{sec:hansen}

Recently, \citet{Ha04} derived a similar but stricter inequality
$\gamma\ge1+\beta$, based on the condition that there exists a
physical power-law solution to the spherically symmetric Jeans and
Poisson equations with the EOS, $\rho\propto r^{-p}\langle v_r^2
\rangle^c$. Since $\beta\le1$, his result is stronger than our result.
However, we note that his result is strictly valid only if both $\rho$
and $\langle v_r^2 \rangle$ behave as the pure power law extending
globally to the infinity. In fact, we find that the condition
$\gamma\ge\beta+1$ originates from the boundary condition at infinity
rather than at the center and thus argue that the result should be
understood to refer to the asymptotic density power index at infinity,
not the central density slope. In particular, the supposed piece of
evidence cited by \citet{Ha04} for his inequality (\citealt{LH00}; see
also \citealt{BJ68}) is in fact due to the constraint imposed by
$\gamma\ge2\beta$ on the central slope through the positivity of the
DF (\S~\ref{sec:rel}; see also \citealt{RT84}), since it involves the
case of purely radial motion ($\beta=1$) for which the two limits
coincide ($\gamma\ge2$).

The general integral solution of the Jeans equation with constant $\beta$
can be written by admitting an integration constant $\tilde B$,
\begin{equation}
r^{2\beta}\rho\langle v_r^2\rangle=\tilde B
+\int_{r_0}^r\!dr'\,r'^{2\beta}\rho(r')\,
\left.\frac{d\psi}{dr}\right|_{r=r'}.
\label{eq:vdg}
\end{equation}
The potential gradient (i.e., the local gravitational acceleration)
can be found from the enclosed mass
\begin{equation}
\frac{d\psi}{dr}=-\frac{GM_r}{r^2}\,;\qquad
M_r=M_{r_0}+4\pi\int_{r_0}^r\rho(r')r'^2dr',
\label{eq:emass}
\end{equation}
where the negative sign is due to our choice of sign for $\psi$.
If we assume a strict power-law behavior for the density as $\rho=
Ar^{-\gamma}$, where $A>0$ is constant, then $M_r$ is given by
($\gamma\ne3$)
\begin{equation}
M_r=
M_{r_0}+4\pi A(3-\gamma)^{-1}\left(r^{3-\gamma}-r_0^{3-\gamma}\right).
\label{eq:ma}
\end{equation}
If the power law extends to the center, the result must be valid for
the choice of $r_0=0$. However, then the mass within any finite
radius $M_r$ always diverges for $\gamma\ge3$ even if $M_0$ is finite,
and therefore the model is unphysical. The resulting upper limit
$\gamma<3$ is well established. By substituting
equation~(\ref{eq:emass}) and $\rho=Ar^{-\gamma}$ in
equation~(\ref{eq:vdg}), $\langle v_r^2\rangle$ as a function of $r$
($\gamma\ne3$, $\gamma\ne\beta+1$, and $\gamma\ne2\beta-1$) is found
to be
\begin{eqnarray}
\langle v_r^2\rangle&=&Br^{\gamma-2\beta}
-r^{2-\gamma}\frac{2\pi GA}{(3-\gamma)(\beta-\gamma+1)}
\nonumber\\&+&\frac{1}{r}
\left[\frac{4\pi GAr_0^{3-\gamma}}{(3-\gamma)(2\beta-\gamma-1)}
-\frac{GM_{r_0}}{2\beta-\gamma-1}\right],
\label{eq:vel}
\end{eqnarray}
where $B$ is an integration constant to be determined from the
boundary condition. Here if we assume strictly scale-free behavior
for $\rho$, equation~(\ref{eq:vel}) is valid everywhere extending from
$r=0$ to $r=\infty$. With $r_0=0$ and $M_0=0$, we can show that the
condition for $\langle v_r^2\rangle$ to be nonnegative everywhere
leads to the inequality $\gamma>\beta+1$, as found by \citet{Ha04},
and $B\ge0$. In addition, the self-similarity implies strict power-law
behavior for $\langle v_r^2\rangle$ as well. Since $A>0$, this can
only be obtained with the choice of $B=0$. In fact, the choice of
$B=0$ can independently be deduced from the boundary condition at
infinity. That is, $\langle v_r^2\rangle$ is nondivergent for a
finite potential ($\gamma<2$), or if the potential diverges
($\gamma>2$), the velocity dispersion cannot diverge faster than the
potential does.

However, if we relax the assumption that $\rho$ is strictly scale-free
everywhere and replace it with $\rho$ being locally well approximated
by the power law near the center, equation~(\ref{eq:vel}) is valid
\emph{only} for the region where $\rho\approx Ar^{-\gamma}$, so
the condition that $\langle v_r^2\rangle$ is nonnegative only needs
to be checked for this regime. Provided that the power law provides a
good approximation to the behavior of $\rho$ near the center and that we
limit our attention to a self-consistent density-potential system, it is
reasonable to choose $r_0=0$ and $M_0=0$. Then for $\gamma>2\beta$,
with any positive constant $B$, equation~(\ref{eq:vel}) returns the
correct behavior of the velocity dispersion near the center, although
its validity does not extend to infinity. For $\gamma>\beta+1$, the
velocity dispersion ($r_0=M_0=0$) is given by
\begin{equation}
\langle v_r^2\rangle=r^{2-\gamma}\left[
\frac{2\pi GA}{(3-\gamma)(\gamma-\beta-1)}+Br^{2(\gamma-\beta-1)}\right].
\end{equation}
Near the center, we find that $\langle v_r^2\rangle\sim
A'r^{2-\gamma}$, where $A'$ is a positive constant. This is just the
approximate local power-law behavior of the velocity dispersion near
the center (valid locally for any $B$). In addition, if $\beta+1<
\gamma\le2$, the central velocity dispersion is finite. Although
$\langle v_r^2\rangle$ diverges at the center if $\gamma>2\ge\beta+1$,
this behavior can be physical because the self-consistent potential
well depth for this case is also infinite. On the other hand, if
$\gamma<\beta+1$, the same velocity dispersion can be written as
\begin{equation}
\langle v_r^2\rangle=r^{\gamma-2\beta}\left[
B-r^{2(\beta+1-\gamma)}\frac{2\pi GA}{(3-\gamma)(\beta+1-\gamma)}\right].
\label{eq:vn}
\end{equation}
In other words, since $\gamma-2\beta<2-\gamma$ if $\gamma<\beta+1$,
the leading term for the velocity dispersion near the center is given
by $\langle v_r^2\rangle\sim Br^{\gamma-2\beta}$ provided that
$B\ne0$. While not strictly scale-free, the local behavior of the
velocity dispersion near the center can still be approximated as a
power law, and furthermore, as long as $B>0$ and $0\le\gamma-2\beta
<2-\gamma$, it is positive and finite. While we note that for a
sufficiently large $r$, 
equation~(\ref{eq:vn}) eventually becomes negative, since the
$r^{2-\gamma}$ term becomes dominant as $r\rightarrow\infty$, this
does not restrict the \emph{central} power index for density, provided
that the behavior of $\rho$ starts to deviate from $\sim Ar^{-\gamma}$
(toward the steeper falloff) at smaller $r$ than the value at which
$\langle v_r^2\rangle=0$ in equation~(\ref{eq:vn}).

Let us consider the explicit example of the \citet{He90} model, which
has a $r^{-1}$ density cusp (i.e., $\gamma=1$). The radial velocity
dispersion of the constant-$\beta$ model is \citep[see, e.g.,][]
{BD02,EA05}
\begin{equation}
\langle v_r^2\rangle=\frac{GM}{(5-2\beta)}\frac{r}{(r+a)^2}\
\mbox{}_2F_1\left(1,5;6-2\beta;\frac{a}{r+a}\right),
\end{equation}
where the potential is given by $\psi=GM/(r+a)$. It is straightforward
to show that, for $0<\beta\le\case12$, $\langle v_r^2\rangle$ is
everywhere positive finite with its leading order behavior near the
center given by $\sim r^{1-2\beta}$ \citep[if $\beta=\case12$, then
$\langle v_r^2\rangle=\psi/4$; see also][]{EA05}. On the other hand,
if $\beta<0$, the leading order for $\langle v_r^2\rangle$ near the
center is found to be $\sim r$ with a positive coefficient. For the
isotropic case ($\beta=0$), the elementary functional expression for
the velocity dispersion is given in equation~(10) of \citet{He90},
whose leading order behavior is found to be $\sim r\ln{r^{-1}}$
\citep[see, e.g., eq.~11 of][]{He90}. 

As another example, we consider the dark matter profile proposed by
\citet{DM05}. They solved the spherically symmetric Jeans and Poisson
equations with the same EOS as in \citet{Ha04}. However, they found a
family of ``realistic'' models with a finite mass and an infinite
extent. With $\rho\propto r^{-p}\langle v_r^2\rangle^{3/2}$, the inner
density cusp of their models is given by $\gamma=(7+10\beta_0)/9$,
where $\beta_0$ is the anisotropy parameter at the center. It is clear
that for all members of these models $\gamma<\beta_0+1$, since
$\beta_0\le1<2$, thus violating the inequality $\gamma\ge\beta+1$. On
the other hand, our result $\gamma\ge2\beta$ indicates that they are
physical if and only if $\beta_0\le\case78$. In fact, $\langle v_r^2
\rangle$ near the center for this family behaves as $\sim
r^{(7-8\beta_0)/9}$, and thus the limit $\gamma\ge2\beta$ is
equivalent to the condition that the central velocity dispersion is
finite.

A similar analysis of equation~(\ref{eq:vel}) can be applied to
discover the asymptotic behavior of the velocity dispersion at
infinity. Suppose that $\rho$ asymptotically approaches a power law
for a sufficiently large $r$. Then for the same range,
equation~(\ref{eq:vel}) is a valid expression for $\langle v_r^2
\rangle$, provided that the power law $\sim A\rho^{-\gamma}$ describes
the asymptotic behavior of $\rho$ and $M_{r_0}$ is the mass within
$r_0$. Here $B=0$ from the boundary condition at infinity. If the
total mass is finite ($\gamma>3>\beta+1>2\beta-1$), we can simply set
$r_0=\infty$ and $M_\infty=M$, where $M$ is the total mass. Then since
$-1>2-\gamma$, the asymptotic behavior of the velocity dispersion is
given by $\langle v_r^2 \rangle\sim (\gamma+1-2\beta)^{-1}GM/r$, which
is consistent with Keplerian falloff. For an infinite-mass system, we
find that $2-\gamma>-1$, and thus the leading term of $\langle v_r^2
\rangle$ for $r\rightarrow\infty$ is $\sim A'r^{2-\gamma}$. Here $A'$
is a positive constant if $\gamma>\beta+1$, whereas it is a negative
constant if $\gamma<\beta+1$. In other words, from the condition of
nonnegativity of the velocity dispersion toward infinity, we actually
recover the inequality of \citet{Ha04}, $\gamma\ge\beta+1$, although
here $\gamma$ and $\beta$ are the asymptotic density power index and
the anisotropy parameter at \emph{infinity}. Although the velocity
dispersion diverges as $r\rightarrow\infty$ if $\gamma<2$, the
behavior may be acceptable because the potential also diverges with
the same power index, so the system is still bounded.


\section{Central Black Hole}
\label{sec:tim}

We have found one further example of a cusp slope--anisotropy
relationship in the literature. \citet{Wh81} found a
relation\footnote{The inequality given in \citet{Wh81} or
\citet{BMZ} did not include the case of equality because of the
specific form of the scale-free DF. The general result actually
extends to include the case of equality through the transition
of the power-law distribution to the Dirac delta distribution.}
by studying the special case of scale-free densities in scale-free
potentials, namely
\begin{equation}
\gamma\ge\frac{\delta}{2}+\beta(2-\delta),
\label{eq:ine}
\end{equation}
where $\delta$ is the central power-law index for the potential (i.e.,
$\psi\sim r^{-\delta}$), which may be externally imposed. Note the
change of the notation from \citet{Wh81}. The form of the limit given
in equation~(\ref{eq:ine}) is actually that of \citet{BMZ}, who
performed a similar study to \citet{Wh81} but allowed for flattening.
This result was derived from a specific functional form for the DF, and
in particular assumed that $f_E(E)$ is scale-free. We note that,
since $\beta\le1$, the inequality $\gamma\ge2\beta$ is automatically
satisfied if $\gamma\ge2$. For $\gamma<2$, the self-consistent
potential-density has a finite central potential, so
$\delta=0$. In this case, equation~(\ref{eq:ine}) reduces to
$\gamma\ge2\beta$. However, in the presence of a central black hole
($\delta=1$), equation~(\ref{eq:ine}) provides us with a different
limit $\gamma\ge\beta+\case12$, which is stricter than $\gamma\ge2\beta$
if $\beta<\case12$.

Here we note that this limit can, without assuming that the density
or DF is scale-free, be derived from the
nonnegativity of the DF for a massless tracer population in the
Keplerian potential of a central point mass. Let us suppose that the
DF for a massless tracer population is given by
equation~(\ref{eq:ansatz}). (The results can then be extended to more
general DFs using the identical argument of \S~\ref{sec:laur}.) Next,
we assume that these tracers are subject to the potential $\psi=GM/r$
of a point mass at the center.  Then the number density of the tracer
population $n$ can be found from integration of the DF over
velocity space as in equation~(\ref{eq:den});
\begin{eqnarray}
n&=&\frac{D_\beta}{r^{2\beta}}
\int_0^{GM/r}\!\left(\frac{GM}{r}-E\right)^{1/2-\beta}f_E(E)\,dE
\nonumber\\&=&\frac{D_\beta}{r^{\beta+1/2}}
\int_0^{GM/r}\!(GM-rE)^{1/2-\beta}f_E(E)\,dE.
\label{eq:epd}
\end{eqnarray}
For $\beta<\case12$, we find that
\begin{equation}
\frac{d}{dr}\left(r^{\beta+1/2}n\right)=-\tilde D_\beta
\!\int_0^{GM/r}\!\frac{Ef_E(E)\,dE}{(GM-rE)^{1/2+\beta}}
\le0
\end{equation}
for any nonnegative DF. Consequently,
\begin{equation}
\frac{d}{dr}\left(r^{\beta+1/2}n\right)=
r^{\beta-1/2}n\left(\beta+\frac{1}{2}+\frac{d\ln{n}}{d\ln{r}}\right)\le0
\end{equation}
or, equivalently,
\begin{equation}
-\frac{d\ln{n}}{d\ln{r}}\ge\beta+\frac{1}{2}
\label{eq:bh}
\end{equation}
which is the desired result. In fact, if the external potential is
replaced by $\psi=C/r^\delta$ in equation~(\ref{eq:epd}), it is easy
to see that equation~(\ref{eq:ine}) simply follows from an
essentially identical argument with $\gamma=-(d\ln{n}/d\ln{r})$ being
the central power index for the number density of the tracer
population.

In fact, the result can still be obtained without assuming the
specific form of (scale-free) potential. That is, provided only that
$\delta=-\lim_{r\rightarrow0}(d\ln\psi/d\ln{r})$ and $0\le\delta\le1$,
we find from equation~(\ref{eq:den}) that
\begin{equation}
n=D_\beta\frac{\psi^{1/2-\beta}}{r^{2\beta}}
\int_0^\psi\!\left(1-\frac{E}{\psi}\right)^{1/2-\beta}f_E(E)dE,
\end{equation}
and thus for $\beta<\case12$, we have
\begin{equation}
\frac{d}{d\psi}\left(nr^{2\beta}\psi^{\beta-1/2}\right)=
\tilde D_\beta
\int_0^\psi\!\frac{E f_E(E)\,dE}{\psi^2[1-(E/\psi)]^{1/2+\beta}}\ge0.
\end{equation}
However, of course,
\begin{eqnarray}\lefteqn{
\frac{d}{d\psi}\left(nr^{2\beta}\psi^{\beta-1/2}\right)=
\frac{nr^{2\beta-1}\psi^{\beta-1/2}}{d\psi/dr}
\frac{d}{d\ln r}\ln\left(nr^{2\beta}\psi^{\beta-1/2}\right)
}\nonumber\\&&=
\frac{nr^{2\beta}\psi^{\beta-3/2}}{d\ln\psi/d\ln r}\left[
\frac{d\ln n}{d\ln r}+2\beta+\left(\beta-\frac{1}{2}\right)
\frac{d\ln\psi}{d\ln r}\right]\ge0,
\end{eqnarray}
and by taking the limit $r\rightarrow0$,
\begin{equation}
\gamma\ge2\beta+\left(\frac{1}{2}-\beta\right)\delta.
\label{eq:bhl}
\end{equation}
Here note that $\lim_{r\rightarrow0}(d\ln\psi/d\ln r)=-\delta\le0$.
This also indicates that the result is still valid even if the
self-gravity of the tracer population is appreciable, as long as the
potential (that may be decoupled from the density) is divergent at the
center.

If $\beta=\case12$, the limit given in equation~(\ref{eq:ine}) is, in
fact, identical to $\gamma\ge2\beta=1$. Since the derivation of the
limit in \S~\ref{sec:rel} for $\beta=\case12$ does not use the assumption
of self-consistency, it is still applicable here.  Therefore, the
limits in equations~(\ref{eq:bh}) and (\ref{eq:bhl}) can be extended to
$\beta\le\case12$. For $\case12<\beta\le1$, the limit $\gamma\ge2\beta$ is
actually stronger than equation~(\ref{eq:ine}) (note that
$0\le\delta\le1$). However, with a centrally diverging potential
(i.e., $\psi_0=\infty$), the proof given in \S~\ref{sec:simon} is not
directly applicable. Instead, from equation~(\ref{eq:lim}), we now
find that $\lim_{r\rightarrow0}k$, where
$k=nr^{2\beta}\psi^{\beta-1/2}$ either is finite and nonzero or
diverges to positive infinity.  Following exactly the same argument as
in \S~\ref{sec:simon} applied to $k=nr^{2\beta}\psi^{\beta-1/2}$
instead of $h=\rho r^{2\beta}$, we have
\begin{equation}
\lim_{r\rightarrow0}\frac{d\ln k}{d\ln r}
=\lim_{r\rightarrow0}\frac{d\ln n}{d\ln r}+2\beta
+\left(\beta-\frac{1}{2}\right)\
\lim_{r\rightarrow0}\frac{d\ln\psi}{d\ln r}\le0,
\end{equation}
which translates to equation~(\ref{eq:bhl}), and thus
we can extend the limit of equation~(\ref{eq:bhl}) to $\beta>\case12$ as well.

The limit $\gamma\ge\beta+\case12$ for $\delta=1$ indicates
that a spherical isotropic system subject to a Keplerian potential
should possess a central density cusp at least as steep as
$r^{-1/2}$. Similarly, if an isotropic stellar system is subject to a
divergent dark matter potential due to a cusped profile with a slope
steeper than that of the isothermal cusp, the stellar system should
also be cusped with its cusp slope constrained to be
$\gamma_\star\ge(\gamma_\mathrm{DM}/2)-1$, where $\gamma_\star$ is the
cusp index for the stellar system and $\gamma_\mathrm{DM}\ge2$ is that
of the dark matter profile. Of course, if $\gamma_\mathrm{DM}<2$, the
central potential is finite provided that there is no other source of
divergent potential, and thus the limit simply reduces to
$\gamma_\star\ge0$. On the other hand, if the system were mildly
radially anisotropic (near the center), that is to say
$\beta\approx\case12$, the limit for the supportable cusp slope would be
steeper, much like $\gamma\ga1$.

\section{Conclusions}

We have proved, for a very wide class of steady-state gravitating
system, a theorem constraining the central cusp slope of the density
profile $\gamma$ (eq.~\ref{eq:gamma}) and the central velocity
anisotropy $\beta$. Specifically, the inequality $\gamma\ge2\beta$ is
a necessary condition for the nonnegativity of the distribution
function (DF). If there is a divergent external potential,
decoupled from the density profile, then the inequality generalizes
to $\gamma\ge2\beta-(\beta-\case12)\delta$.
Here the external potential diverges as $\sim r^{-\delta}$ at the
center. Finally, if the external potential is due to a central black
hole, it reduces to $\gamma\ge\beta+\case12$. We expect our
result to be useful in the study of dense stellar systems, or in the
building of extreme stellar dynamical models.

As most N-body simulations predict only very modest anisotropies
($\beta\approx0$) in the very center, the application of our result
does not directly constrain the central density profile
($\gamma\ga0$). While the inequality derived by \citet{Ha04}, namely,
$1+\beta\le\gamma<3$, appears to be stronger than our result, his
lower limit is only strictly applicable to the scale-free power-law
density profile of infinite extent. It appears that $\gamma\ge\beta+1$
actually constrains the asymptotic behavior of the density power index
and the anisotropy parameter at infinity rather than at the center.

\acknowledgments We thank P.~Tim de~Zeeuw for stimulating discussions
and helpful suggestions.  We appreciate the generous suggestions made
by the anonymous referee to improve the paper. We are grateful to
Simon~D.~M. White for noticing a flaw in an earlier version. We
acknowledge S. Hansen for pointing us to the results of \citet{DM05}
and others.

\begin{appendix}

Here we provide a more detailed argument for the generality of the
theorem than that given in \S~\ref{sec:laur}. While we do not claim
that the following argument strictly adheres to the high standard of
the pure mathematician, we hope that it indicates the generality of
the result.

First, let us suppose that a DF $f(E,L)$ can be written as
\begin{equation}
f(E,L)=L^{-2\beta_0}\left[f_0(E)+f_1(E,L)\right],
\label{eq:dfg}
\end{equation}
where $f_0(E)$ is a function of $E$ alone, whereas
$f_1(E,L)$ is a continuous function that satisfies
\begin{equation}
f_1(E,L=0)=0,
\end{equation}
which further implies that $f_0(E)\ge0$ for all accessible values of
$E$ from the nonnegativity of the DF. Then
\begin{eqnarray}
\label{eq:gd}
\rho&=&2\pi\iint
L^{-2\beta_0}\left[f_0(E)+f_1(E,L)\right]
v_\mathrm{T}dv_\mathrm{T}dv_r
\nonumber\\
&=&\frac{D_{\beta_0}}{r^{2\beta_0}}
\int_0^\psi\!dE\,(\psi\!-\!E)^{1/2-\beta_0}f_0(E)
+\frac{4\pi}{r^{2\beta_0}}
\iint\!dv_\mathrm{T}dv_r\,v_\mathrm{T}^{1-2\beta_0}
f_1\left(\psi\!-\!\frac{v_\mathrm{T}^2+v_r^2}{2},rv_\mathrm{T}\right)
\\
\rho\langle v_r^2\rangle&=&2\pi\iint
v_r^2L^{-2\beta_0}\left[f_0(E)+f_1(E,L)\right]
v_\mathrm{T}dv_\mathrm{T}dv_r
\nonumber\\
&=&\frac{2D_{\beta_0}}{r^{2\beta_0}(3\!-\!2\beta_0)}
\int_0^\psi\!dE(\psi\!-\!E)^{3/2-\beta_0}f_0(E)
+\frac{4\pi}{r^{2\beta_0}}
\iint\!dv_\mathrm{T}dv_r\,v_\mathrm{T}^{1-2\beta_0}v_r^2
f_1\left(\psi\!-\!\frac{v_\mathrm{T}^2+v_r^2}{2},rv_\mathrm{T}\right)
\qquad\\
\rho\langle v_\mathrm{T}^2\rangle&=&2\pi\iint
v_\mathrm{T}^2L^{-2\beta_0}\left[f_0(E)+f_1(E,L)\right]
v_\mathrm{T}dv_\mathrm{T}dv_r
\nonumber\\
&=&\frac{4D_{\beta_0}(1\!-\!\beta_0)}{r^{2\beta_0}(3\!-\!2\beta_0)}
\int_0^\psi\!dE(\psi\!-\!E)^{3/2-\beta_0}f_0(E)
+\frac{4\pi}{r^{2\beta_0}}
\iint\!dv_\mathrm{T}dv_r\,v_\mathrm{T}^{3-2\beta_0}
f_1\left(\psi-\frac{v_\mathrm{T}^2+v_r^2}{2},rv_\mathrm{T}\right).
\end{eqnarray}
Taking the limit $r\rightarrow0$, the velocity moment integrals of the
$f_1$ term vanishes, provided that the domain over which it is
integrated is bounded and that $f_1$ is sufficiently
well behaved. Note then that the anisotropy parameter at the center
for this model is found to be
\begin{equation}
\lim_{r\rightarrow0}\beta=1-
\lim_{r\rightarrow0}\frac{\rho\langle v_\mathrm{T}^2\rangle}
{2\rho\langle v_r^2\rangle}=1-(1-\beta_0)=\beta_0.
\label{eq:betag}
\end{equation}

Next, suppose that $h_0=r^{2\beta_0}\rho_0$ and $\rho_0$ is the
density profile built by the DF, $f(E,L)=L^{-2\beta_0}f_0(E)$.
Then the proof given in \S~\ref{sec:con} indicates that
$\gamma_0\ge2\beta_0$, where
\begin{equation}
\gamma_0=-\lim_{r\rightarrow0}\frac{d\ln\rho_0}{d\ln r}.
\end{equation}
Now, since $\lim_{r\rightarrow0}(d\ln h_0/d\ln r)=2\beta_0-\gamma_0\le
0$, we have $\lim_{r\rightarrow0}h_0\ne0$, that is,
$\lim_{r\rightarrow0}h_0$ is nonzero finite or diverges.
Subsequently, from equation~(\ref{eq:gd}),
\begin{equation}
\lim_{r\rightarrow0}\left(r^{2\beta_0}\rho-h_0\right)=
\lim_{r\rightarrow0}\left(\frac{r^{2\beta_0}\rho}{h_0}-1\right)h_0=0,
\end{equation}
but $\lim_{r\rightarrow0}h_0\ne0$, and therefore we find that
\begin{equation}
\lim_{r\rightarrow0}\frac{r^{2\beta_0}\rho}{h_0}=1.
\label{eq:limrt}
\end{equation}
If $\lim_{r\rightarrow0}h_0$ is finite (i.e. $\gamma_0=2\beta_0$),
then equation~(\ref{eq:limrt}) implies that
\begin{equation}
0<\lim_{r\rightarrow0}r^{2\beta_0}\rho=\lim_{r\rightarrow0}h_0<\infty,
\label{eq:dfi}
\end{equation}
so $\lim_{r\rightarrow0}r^{2\beta_0}\rho$ is also finite,
that is, $\rho\sim r^{-2\beta_0}$.
On the other hand, if $\lim_{r\rightarrow0}h_0=\infty$,
l'H\^opital's rule indicates that
\begin{equation}
\lim_{r\rightarrow0}\frac{r^{2\beta_0}\rho}{h_0}=
\lim_{r\rightarrow0}\frac{d(r^{2\beta_0}\rho)/dr}{dh_0/dr}=
\lim_{r\rightarrow0}\frac{r^{2\beta_0}\rho}{h_0}
\frac{d\ln(r^{2\beta_0}\rho)/d\ln r}{d\ln h_0/d\ln r}=
\frac{1}{2\beta_0-\gamma_0}
\left(2\beta_0-\lim_{r\rightarrow0}\frac{d\ln\rho}{d\ln r}\right)=1.
\label{eq:din}
\end{equation}
In other words, combining the results in equations~(\ref{eq:dfi})
and (\ref{eq:din}), we have established that
\begin{equation}
\lim_{r\rightarrow0}\frac{d\ln\rho}{d\ln r}=\gamma_0=
\lim_{r\rightarrow0}\frac{d\ln\rho_0}{d\ln r},
\label{eq:cuspg}
\end{equation}
where $\rho$ is the density profile built by the DF of
equation~(\ref{eq:dfg}) and $\rho_0$ by $f(E,L)=L^{-2\beta_0}f_0(E)$.
Finally, taking equations~(\ref{eq:betag}) and (\ref{eq:cuspg})
together, we have established that the theorem $\gamma_0>2\beta_0$
extends to a more general class of DFs of the form of
equation~(\ref{eq:dfg}).

How general is the form of the DF in equation~(\ref{eq:dfg})? We argue
that it is almost always possible to reduce most well-behaving DFs to
the form of equation~(\ref{eq:dfg}). That is, for a general DF of a
spherically symmetric system in equilibrium, the reduction is possible
if there exists $\alpha<2$ such that
\begin{equation}
f_E(E)=\lim_{L\rightarrow0}L^{\alpha}f(E,L),
\end{equation}
where $f_E(E)$ should be finite and nonzero for values of $E$ at
least in some nonempty subset of all the accessible values of
$E\in[0,\psi_0]$. Then the original DF can be written in a form of
equation~(\ref{eq:dfg}) as
\begin{equation}
f(E,L)=L^{-\alpha}\left[f_E(E)+f_1(E,L)\right]\,;\qquad
f_1(E,L)=L^{\alpha}f(E,L)-f_E(E).
\end{equation}
and it is obvious to show that
\begin{equation}
f_1(E,L=0)=\lim_{L\rightarrow0}L^{\alpha}f(E,L)-f_E(E)=0.
\end{equation}
For example, for the DF of the form given by \citet{Cu91},
\begin{equation}
f(E,L)=L^{-2\beta_0}f_0(Q)\,;\qquad
Q\equiv E-\frac{L^2}{2r_\mathrm{a}^2},
\end{equation}
the reduction is given by
\begin{equation}
f(E,L)=L^{-2\beta_0}\left[f_0(E)+f_1(E,L)\right]\,;\qquad
f_1(E,L)=f_0(Q)-f_0(E).
\end{equation}
\end{appendix}
\label{end}

\end{document}